\documentclass[runningheads]{llncs}
\usepackage[T1]{fontenc}
\usepackage{graphicx}
\usepackage{csquotes}
\MakeOuterQuote{"}
\usepackage{tikz}
\usetikzlibrary{positioning, arrows.meta, shapes, fit, decorations.pathreplacing}
\usepackage{amsmath}
\usepackage{algpseudocode}
\usepackage{algorithm}
\usepackage{url}

\usepackage{breakurl}
\usepackage[breaklinks]{hyperref}
\usepackage{amsfonts}
\usepackage{booktabs}
\usepackage{multirow}
\usepackage{chngpage}
\usepackage{adjustbox}
\usepackage{rotating}
\usepackage{color}

\newif\iflong
\longtrue

\begin{document}
\title{Kintsugi: Decentralized E2EE Key Recovery}
\author{Emilie Ma\inst{1}\orcidID{0009-0005-3322-0805} \and
Martin Kleppmann\inst{2}\orcidID{0000-0001-7252-6958}}
\authorrunning{E. Ma \and M. Kleppmann.}
\institute{University of British Columbia, British Columbia, Canada\\
\email{contact@emilie.ma}\\
\and
University of Cambridge, Cambridge, United Kingdom\\
\email{martin.kleppmann@cst.cam.ac.uk}}
\maketitle              % typeset the header of the contribution
\begin{abstract}

Kintsugi is a protocol for key recovery, allowing a user to regain access to end-to-end encrypted data after they have lost their device, but still have their (potentially low-entropy) password. Existing E2EE key recovery methods, such as those deployed by Signal and WhatsApp, centralize trust by relying on servers administered by a single provider.
% This can be problematic for applications requiring metadata privacy or wanting to avoid a single party controlling user identities, for example.
Kintsugi is decentralized, distributing trust over multiple recovery nodes, which could be servers run by independent parties, or end user devices in a peer-to-peer setting. To recover a user's keys, a threshold $t+1$ of recovery nodes must assist the user in decrypting a shared backup. Kintsugi is password-authenticated and protects against offline brute-force password guessing without requiring any specialized secure hardware.
% We make use of proactive, dynamic-committee secret sharing in order to enable users to update recovery nodes and nodes to regularly regenerate their shares without changing the joint secret itself. 
\iflong
Kintsugi can tolerate up to $t$ honest-but-curious colluding recovery nodes, as well as $n - t - 1$ offline nodes, and operates safely in an asynchronous network model where messages can be arbitrarily delayed.
\fi

\keywords{Decentralized key recovery \and End-to-end encryption.}
\end{abstract}

\section{Introduction}

As end-to-end encrypted (E2EE) services gain popularity, an important problem is account or key recovery. \iflong With non-E2EE services, a user who loses their device or switches to a new one can simply log in on a different machine with their username and password. In an E2EE setting, this is problematic: the password may lack sufficient entropy, so it cannot securely seed a key for decrypting a backup of the user's data. \fi Common recovery mechanisms for E2EE platforms include user-selected recovery passwords \cite{Whatsapp_Recovery}, recovery codes \cite{Whatsapp_RecoveryCodes,MEGA_Recovery_Key,LastPass}, short PINs with hardware-enforced guess limits \cite{Connell_Signal_Key_Recovery}, local copies of recovery files \cite{1Password}, or a designated recovery contact \cite{Apple_Recovery_Contact,1Password,Preveil_Whitepaper}. However, a key issue with existing recovery methods is the trend towards centralization: users cannot verify that the central service implements recovery securely or reliably. Existing mechanisms have associated tradeoffs: 

\begin{itemize}
\item Signal's Secure Value Recovery \cite{Connell_Signal_Key_Recovery} and WhatsApp's E2EE backups \cite{Whatsapp_Recovery} require the user to remember a four-digit PIN and rely on trusted hardware, like hardware security modules (HSMs), to limit PIN guesses. \iflong In the case of Signal, multiple hardware elements are distributed across several locations and cloud providers. Nevertheless, the hardware on both platforms remains a centralized system, operated by a single party that must be trusted to correctly manage the key recovery infrastructure. \else The hardware is thus a centralized system operated by a single party that must be trusted to correctly manage the key recovery infrastructure.\fi

\iflong
\item Apple iCloud allows users to designate a single recovery contact \cite{Apple_Recovery_Contact}. If this contact is untrustworthy and has physical access to one of the user's devices, they are able to take over the user's account. 
\else
\item Apple iCloud allows users to designate a single recovery contact \cite{Apple_Recovery_Contact}. If this contact is untrustworthy, they are able to take over the user's account. 
\fi

\item \iflong Recovery codes used by services like MEGA \cite{MEGA_Recovery_Key} or LastPass \cite{LastPass}, and Bitcoin wallet recovery seed phrases consisting of twelve random words \cite{BIP39}, avoid trusting any other parties. However, their high-entropy nature makes them impractical to remember, and paper backups are prone to being lost. Digital recovery files are at risk of device loss or failure or improper storage in an unencrypted cloud service, undermining the system's E2EE properties \cite{Blessing_Hugenroth_Anderson_Beresford_2024}. \else Recovery codes \cite{MEGA_Recovery_Key,LastPass} or recovery seed phrases \cite{BIP39} avoid trusting any other parties, but their high-entropy nature makes them impractical to remember. \fi

\item Applying Shamir Secret Sharing \cite{Shamir_1979} to split a recovery key across multiple contacts raises the issue of authentication: when a contact receives a request to participate in secret reconstruction, the contact needs to decide if the request is genuine. \iflong This is susceptible to social engineering, tricking contacts into revealing their secret shares to an adversary. There is also no protection against a threshold of contacts colluding to reconstruct the key directly and no way to change the set of contacts without repeating the secret-sharing process.\else This is susceptible to social engineering and to malicious, colluding contacts. \fi

\end{itemize}

\iflong
Centralized infrastructure is a risk for applications that require metadata privacy (e.g. anonymity networks) or where the infrastructure may be shut down outside of their control (e.g. hosting for government-sanctioned activists), besides requiring users to blindly trust the service. For example, WhatsApp uses a HSM-based Backup Key Vault service to protect against brute-force attacks, but this requires trusting the HSMs, which are under WhatsApp's sole control \cite{Whatsapp_HSM}.

There is also an aspect of cost: Signal, WhatsApp, and Apple iCloud all rely on HSMs to ensure only authenticated users can recover their data. However, these HSMs are expensive and difficult to deploy. For instance, Signal's staging (non-production) Secure Value Recovery implementation costs \$2,100/month to run \cite{Connell_Signal_Key_Recovery}. In turn, this limits decentralization efforts: while volunteers may be happy to help run nodes for a decentralized service on consumer-level hardware (e.g. Tor node operators), it becomes cost-prohibitive to participate if specialized hardware is required. We thus wished to explore a new space in recovery protocols that avoids any reliance on centralized infrastructure or specialized hardware.
\else
Centralized infrastructure is a risk for applications that require metadata privacy (e.g. anonymity networks) or where the infrastructure may be shut down outside of their control (e.g. hosting for government-sanctioned activists). Users additionally cannot verify that the central service implements recovery securely or reliably: users must blindly trust it. There is also an aspect of cost for HSM-based mechanisms, as HSMs are expensive and complex to deploy. For instance, Signal's staging (non-production) implementation costs \$2,100/month to run \cite{Connell_Signal_Key_Recovery}. We thus wished to explore new recovery protocols that avoid relying on centralized infrastructure or specialized hardware.
\fi

In this paper, we propose Kintsugi, a decentralized key recovery protocol based on a peer-to-peer network of recovery nodes. Nodes may be servers operated by different parties, or end-user devices of contacts, or a mixture of the two. To recover their secret key on a new device, the user must keep track of and provide a password. Then, the device communicates with at least $t+1$ recovery nodes, where $t$ is the reconstruction threshold chosen by the user when they set up their recovery contacts. The recovery nodes do not need to check that requests are authentic; they only need to rate-limit requests to prevent online brute-force attacks on the user's password. \iflong Kintsugi is capable of handling up to $t$ honest-but-curious nodes who participate in the protocol correctly but may collude in trying to compromise the user's secrets, and $n - t - 1$ offline nodes, where $n$ is the total number of recovery nodes for this user. To obtain the user's secret key, $t+1$ recovery nodes would need to collude and additionally mount an offline brute-force attack on the user's password. Users can update their recovery nodes at any time, and even if some recovery nodes are compromised over time, the user's account remains secure thanks to secret refreshing, which regularly regenerates the shares of secrets held by the nodes without changing the joint secret itself. Kintsugi can also operate in an asynchronous network model, where messages may be arbitrarily delayed. 
\else
Users can update their recovery nodes at any time, and even if recovery nodes are compromised over time, the user's keypair remains secure thanks to secret refreshing.
\fi

\section{Threat Model}

We assume that all recovery nodes correctly follow the protocol (i.e. are not Byzantine). Byzantine fault-tolerance is not yet implemented in our prototype, but could be added based on prior work \cite{Yurek_Xiang_Xia_Miller_2022}. Each user can choose the number of recovery nodes they use, $n$, and the key reconstruction threshold, $t$. Kintsugi can tolerate up to $t$ honest-but-curious recovery nodes, which may collude, attempt to guess the user's password, and do not rate-limit responses, but otherwise correctly follow the protocol, and $n - t - 1$ offline nodes, which do not respond to recovery requests. 

We assume an adversary who can both passively eavesdrop and actively interfere with network traffic, including making recovery requests to recovery nodes. Additionally, the adversary can obtain the secret shares of up to $t$ recovery nodes (the same $t$ honest-but-curious nodes previously mentioned) within any given refresh interval and perform brute-force attacks, although we assume they cannot compute discrete logarithms. Alternatively, we also allow an adversary who obtains more than $t$ secret shares, but who lacks computational resources such that offline brute-force attacks have a negligible success rate.

Finally, we assume an asynchronous network model, where messages may be arbitrarily delayed.

\section{Kintsugi Protocol}

\iflong
Several properties are required for Kintsugi's design.

\begin{itemize}
    \item Recovery of the user's key requires the participation of a group of recovery nodes. An adversary wanting to reconstruct the user's key should require both collusion among at least $t+1$ recovery nodes and brute-force password guessing. We accomplished this via an threshold Oblivious Pseudo-Random Function (OPRF) exchange. See section \ref{preliminaries-oprf} for more details. This OPRF usage was inspired by the OPAQUE protocol \cite{Wood_Bourdrez_Lewi_Krawczyk}.
    \item A threshold of recovery nodes must be involved in the key reconstruction. This requires a secret sharing scheme, like Shamir's Secret Sharing \cite{Shamir_1979}.
    \item Users must be able to change their recovery nodes and the threshold of nodes required to recover the user's key. This must be possible at any time, not requiring a delay until the start of a next epoch. Former recovery nodes should also not be able to participate in the reconstruction of the user's key. This requires a \emph{dynamic, proactive} secret sharing scheme: the set of recovery nodes can be updated (dynamic committee) and the shares held by recovery nodes can be renewed while keeping the joint secret the same (proactive refresh). Using a proactive secret sharing scheme also guards against shares being compromised over time, since old shares will be invalidated on each refresh.
    \item Kintsugi should continue to correctly operate in an asynchronous network model where messages can be arbitrarily delayed, as we reasonably expect recovery nodes to go offline (e.g. maintenance, server outage, network interruption). Synchronous protocols become unsafe in an asynchronous network, because the nodes whose messages are delayed may be ejected, causing fault-tolerance against honest-but-curious or genuinely offline nodes to decrease.
\end{itemize}
\fi

Kintsugi consists of three protocols: user registration, key recovery, and changing recovery nodes. \iflong A prototype implementation, built with Rust, Tauri, and libp2p, is available as open source\footnote{\url{https://github.com/kewbish/kintsugi}}. \else A Rust implementation is being developed as open source\footnote{\url{https://github.com/kewbish/kintsugi}}. \fi

\subsection{Threshold OPRF}
\label{preliminaries-oprf}

We combine an Oblivious Pseudo-Random Function (OPRF) with Shamir secret sharing (SSS) \cite{Shamir_1979} to obtain a threshold OPRF \cite{Jarecki_Kiayias_Krawczyk_Xu_2017}, which is performed once per user registration or key recovery request. This OPRF output is used as an encryption key for the recovery data backup.

Consider an elliptic curve group $\mathbb{E}$ of prime order $q$. We use the Ristretto curve \cite{Ristretto}, a modification of Curve25519 to eliminate cofactors, because of its efficiency and its compatibility with off-the-shelf OPRF libraries. Let $\mathit{pwd}$ be the user's password. Kintsugi's OPRF is a deterministic function $f(P, s)$ where:

\begin{itemize}
    \item $P = \textsc{HashToCurve}(\mathit{pwd}) \in \mathbb{E}$, the user's password hashed to a point on the curve using a scheme like Elligator \cite{Bernstein_Hamburg_Krasnova_Lange_2013}
    \item $s \in \mathbb{Z}_q$ is a secret selected by the user's device during registration, of which each recovery node $j$ holds a SSS share $s_j$
    \item the user inputs $P$ to the function $f$ without knowing $s$, and learns the output $f(P, s) = s \cdot P$\iflong, the scalar product in the elliptic curve group\fi
    \item the recovery node $j$ inputs $s_j$ to the function, and learns neither $P$ nor $s \cdot P$
    \item the output $f(P, s) = s \cdot P$ is computationally indistinguishable from random
\end{itemize}

\iflong
To perform an OPRF evaluation, the user's device first generates the uniformly random secret $s \in \mathbb{Z}_q$ and splits $s$ into SSS shares that are distributed to the recovery nodes. SSS works by defining a threshold $t$ and a polynomial $SSS(x) = s + c_1 x + c_2 x^2 \dots c_t x^t$, where the coefficients $c_i \in \mathbb{Z}_q$ are sampled uniformly randomly. For each user, each recovery node is assigned an integer index $i = 1, 2, 3 \dots$, and its secret share $s_i = SSS(i)$ (i.e. the polynomial evaluated at its index using arithmetic in the field $\mathbb{Z}_q$). Any collection of $t + 1$ points can then be interpolated together to recover the secret constant term, which is $s = SSS(0)$.
\fi

\iflong
\else
\vspace{-6ex}
\fi
\begin{figure}
    \centering
    \caption{Threshold OPRF evaluation after secret share setup.}
    \label{fig-threshold-oprf-explanation}
\begin{tikzpicture}[node distance=1.2cm, font=\small]
    \node (user) [align=right, anchor=east] {User};
    \node (contact) [align=right, below=of user.east, anchor=east] {Recovery\\Node 1};
    \node (contacti) [align=right, below=of contact.east, anchor=east] {Recovery\\Node $i$};
    \node (dots) [align=right, below=0.6cm of contacti.east, anchor=east] {$\dots$};
    \node (contactt) [align=right, below=0.6cm of dots.east, anchor=east] {Recovery\\Node $t + 1$};
    
    \node[right=10.5cm of user.east] (user_ground) {};
    \node[right=10.5cm of contact.east] (contact_ground) {};
    \node[right=10.5cm of contacti.east] (contacti_ground) {};
    \node[right=10.5cm of contactt.east] (contactt_ground) {};
    
    \draw[->] (user) -- (user_ground);
    \draw[->] (contact) -- (contact_ground);
    \draw[->] (contacti) -- (contacti_ground);
    \draw[->] (contactt) -- (contactt_ground);
    
    \node (user2) [below right=-0.15cm and 0.1cm of user.east] {};
    \node (contact1) [below right=1.2cm and 0.5cm of user2.east] {};
    \draw[->] (user2) -- node[midway, above, sloped, fill=white] {\( r_1 \cdot P \)} (contact1);
    \node (user3) [right=0.5cm of user2.east] {};
    \node (contactj) [below right=2.4cm and 1.2cm+0.5cm+0.1cm of user2.east] {};
    \draw[->] (user3) -- node[midway, above, sloped, fill=white] {\( r_i \cdot P \)} (contactj);
    \node (user4) [right=1.2cm of user2.east] {};
    \node (contactt1) [below right=2.4cm+1.2cm and 1.2cm+0.6cm+1.2cm+0.1cm of user2.east] {};
    \draw[->] (user4) -- node[midway, above, sloped, fill=white] {\( r_{t+1} \cdot P \)} (contactt1);

    \node (contact2) [below right=1.2cm and 3cm of user2.east] {};
    \node (user5) [right=3cm+0.6cm of user2.east] {};
    \draw[opacity=0,->] (contact2) -- node[midway, above, sloped, fill=white,opacity=1] {\( s_1 \cdot r_1 \cdot P \)} (user5);
    \draw[->] (contact2) -- (user5);
    
    \node (contactj2) [below right=2.4cm and -0.6cm of user5.east] {};
    \node (user6) [right=1.2cm-0.6cm of user5.east] {};
    \draw[->] (contactj2) -- node[midway, above, sloped, fill=white] {\( s_i \cdot r_i \cdot P \)} (user6);
    
    \node (contactt2) [below right=3.6cm and -0.4cm of user5.east] {};
    \node (user7) [right=1.8cm-0.3cm of user5.east] {};
    \draw[->] (contactt2) -- node[midway, above, sloped, fill=white] {\( s_{t+1} \cdot r_{t+1} \cdot P \)} (user7);

    \node (user8anchor) [below right=0px and 2cm of user7.east] {};
    \node (user8) [rectangle callout, callout absolute pointer=(user8anchor), draw, fill=white, minimum width=3cm, align=center, below=2px of user8anchor.south, inner sep=6px, text depth=5px] {\vspace{3px} \\ Multiply each\\response by $\lambda_i(0) \cdot r_i^{-1}$ \\ \tikz \draw (0,0) -- (3,0); \vspace{4px} \\ Sum responses\\$= s \cdot P$ };

\end{tikzpicture}
\end{figure}
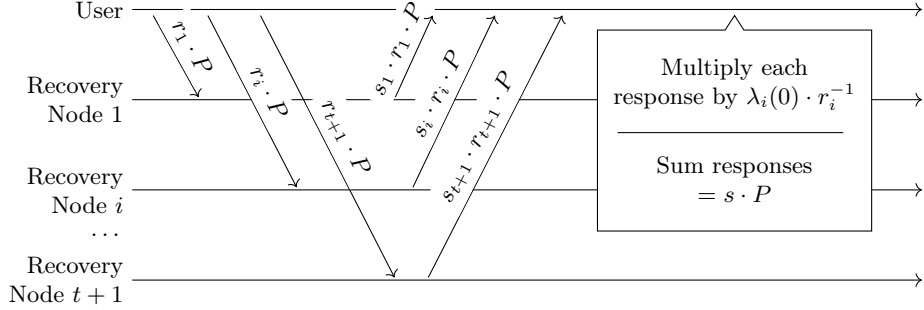

See Figure \ref{fig-threshold-oprf-explanation} for an example OPRF evaluation flow. The user's device chooses and stores a random blinding scalars $r_i \in \mathbb{Z}_q$ to prevent recovery nodes from learning $P$. The user's device sends $r_i \cdot P$ to every recovery node $i$, which multiplies it by $s_i$ and returns the result. After the user's device has received $t + 1$ such results, it multiplies each response $s_i \cdot r_i \cdot P$ by $r_i^{-1}$ to obtain $s_i \cdot P$. \iflong It then applies Lagrange interpolation to find $s \cdot P$. \else It then applies Lagrange interpolation by multiplying by $\lambda_i(0)$, the Lagrange interpolation coefficient for index $i$ and the set of indices of the other recovery nodes, and summing the results to find $s \cdot P$.\fi

\iflong
Let $\lambda_i(x)$ be the Lagrange polynomial:

\begin{align*}
\lambda_i(x) = \prod_{\substack{1 \le m \le t+1 \\ m \neq i}} \frac{x - x_m}{x_i - x_m} &&& \parbox{20em}{where $x_k$ is the index at which the SSS polynomial was evaluated (i.e. the recovery node's index) for the $k$th secret share being interpolated.}
\end{align*}

Then, the Lagrange interpolation to reconstruct $s \cdot P$ is:
\begin{align*}
    & \lambda_1(0) \cdot s_1 \cdot P + \lambda_2(0) \cdot s_2 \cdot P + \cdots + \lambda_{t+1}(0) \cdot s_{t+1} \cdot P \\
    &= (\lambda_1(0) \cdot s_1 + \lambda_2(0) \cdot s_2 + \cdots + \lambda_{t+1}(0) \cdot s_{t+1}) \cdot P\\
    &= s \cdot P \text{  (by Lagrange interpolation of the shares of $s$)}
\end{align*}

This computation is performed in the group $\mathbb{E}$, with $\cdot$ denoting scalar multiplication and $+$ denoting point addition. However, the result is the same as if the secret $s$ had been reconstructed in $\mathbb{Z}_q$ field arithmetic, then multiplied with the point $P$: the interpolation is performed "in the exponent" \cite{Jarecki_Kiayias_Krawczyk_Xu_2017}.
\fi

The output $s \cdot P$ is the encryption key for the recovery data backup, as detailed below. Crucially, no party in this protocol can reconstruct $s$, because each node only returns the scalar multiplication of their share with the user-provided password point, and never the actual share. Even if the point $P$ is chosen by an adversary, they cannot compute $s$, as that would require computing a discrete log.

\iflong
The protocol would be insecure if $P = p \cdot G$, where $G$ is a generator of $\mathbb{E}$ and $p$ is an efficiently computable function of \textit{pwd}. The adversary could send $G$ to the recovery nodes without knowing \textit{pwd}, compute $s \cdot G$ from the responses, and then perform an offline brute-force attack on the password, trying many values $p$ until the correct OPRF output $p \cdot s \cdot G = s \cdot P$ is returned. In contrast, if the discrete logarithm of $P$ is not known, as is the case when using a scheme like Elligator \cite{Bernstein_Hamburg_Krasnova_Lange_2013} to hash $\mathit{pwd}$ to a point on the curve, the adversary must send a separate request to the recovery nodes for each password attempt. This turns an offline brute-force attack into an online one, allowing recovery nodes to enforce rate-limiting on password guesses.
\fi

\subsection{Registration and Key Recovery}

During registration, the user provides a username and a password, and hashes their password into their secret curve point $P$. The user also samples a uniformly random secret $s \in \mathbb{Z}_q$, splits it into SSS shares, and distributes shares to their selected recovery nodes via encrypted and authenticated channels. The OPRF output $s \cdot P$ is then used as the key to encrypt a backup of the user's recovery key, and any other data they wish to restore after a device loss, with an authenticated encryption scheme. This encrypted backup is sent to all recovery nodes, which store it along with their secret share $s_j$. 

When recovering their key, we assume users will not have access to or remember their set of recovery nodes. On registration, we use a distributed hash table (DHT), a decentralized, replicated key-value store, to store a mapping from username to recovery node identifiers.

\iflong
In our implementation, we used the Kademlia DHT \cite{Kademlia} provided by libp2p \cite{Kademlia_Libp2p} to map usernames to a list of the recovery nodes' libp2p addresses. To prevent DHT poisoning or unauthorized updates, users provide their public key when first updating their DHT entry and sign subsequent updates. This approach is not confidential, leaking the identities of each user's chosen recovery nodes. This may be problematic for privacy if the recovery nodes are the user's contacts in a peer-to-peer setting, for example. However, the user would otherwise be required to keep track of their recovery nodes, and we argue that exposing this limited information is an acceptable tradeoff. We also note that Kademlia does not provide any integrity guarantees: this is not an issue given our threat model assumes no Byzantine nodes, but a misbehaving DHT node could otherwise serve fake recovery node information. Future implementations of Kintsugi may benefit from a different, BFT DHT or from storing recovery node information on a central server to simplify the threat model. 
\fi

Users recover their key by looking up their recovery nodes by username in the DHT, inputting their password, and then initiating a threshold OPRF evaluation. This recovery can be performed on a different device than the device used during user registration. The user also downloads their encrypted key backup from one of the recovery nodes. \iflong If the password was correct, they can decrypt the backup with the computed OPRF output $s \cdot P$, and thus recover their data. If decryption fails, this may indicate that the password was incorrect, or that one of the recovery node responses was corrupted or modified by an active network adversary. The only way of trying a new password is to restart the threshold OPRF evaluation with a new $P$. Recovery attempts may be replicated as a transparent audit log across nodes to alert users to potentially fraudulent recovery attacks if high volumes of attempts are made: we leave this as future work.
\else 
If the password was correct, they can decrypt the backup with the OPRF output $s \cdot P$, and thus recover their data. If the password was incorrect, decryption fails. The only way to try a new password is to restart the threshold OPRF evaluation with a new $P$.
\fi

\iflong During recovery, users exchange unauthenticated messages with their recovery nodes. This enables users to initiate contact without knowing anything besides their username and password. \fi Recovery nodes must implement rate-limiting to prevent against online brute-force attacks: for example, rate-limiting recovery requests by IP. This rate-limiting does not depend on HSMs: if at most $t$ recovery nodes fail to rate-limit attempts, any attackers sending forged requests will still need to wait for the slowest recovery node among the $t+1$ reconstruction nodes to return a result before the attacker can check their attempt. \iflong Note that if IP rate-limiting is applied, attackers may be able to use a botnet or rent large blocks of IPs in the cloud to circumvent the rate-limit. However, rate-limiting instead per user may lead to denial-of-service attacks if attackers impersonate a user to use up their quota before they can submit a genuine recovery request. We leave this tradeoff as an area for future exploration.\fi

\iflong
Users are free to choose their reconstruction threshold $t$ and the total number of recovery nodes $n$. In our prototype implementation's UI, we defaulted to $t = 3, n = 5$. Choosing $n \gg t$ is convenient for increased availability if some nodes are offline, but an adversary has more recovery nodes to choose from to potentially compromise and reconstruct $s$.
\fi

\subsection{Dynamic Proactive Secret Sharing}

\iflong
Dynamic proactive secret sharing (DPSS) combines two of the required properties: users must be able to update their set of recovery nodes, and former nodes must not be able to participate in recovering the user's key. DPSS is used to refresh each recovery node's secret share $s_j$ used in the threshold OPRF.
\else
Dynamic proactive secret sharing (DPSS) combines two desirable properties: users can update their set of recovery nodes, and former nodes cannot participate in recovering the user's key. DPSS is used to refresh each recovery node's secret share $s_j$ used in the threshold OPRF.
\fi

We use the high-threshold Honey Badger approach by Yurek et al. \cite{Yurek_Xiang_Xia_Miller_2022}. Honey Badger is an \emph{asynchronous} DPSS protocol, which means that it remains operational in the face of unpredictable network delays\iflong: this fulfills our final requirement for our choice of DPSS\fi. \iflong \else Honey Badger works by generating a new SSS polynomial $SSS'_i(x)$ based on the old secret share and using this refreshed polynomial as the basis for future secret reconstruction. \fi

\iflong
The core idea of Honey Badger is for each node $i$ to generate a new SSS polynomial $SSS'_i(x) = s_i + c_1'x + c_2'x^2 \dots c_{t'}' x^{t'} $ with the new polynomial's constant term being the node's old secret share $s_i$. The node then broadcasts $SSS'_i(j)$ to each other node $j$, which allows node $j$ to interpolate a new node share $s_j'$ that represents the original nodes' shares at index $j$. These new node shares can be further interpolated to recover the original secret $s$. Recall that each $s_i = SSS(i)$ and $s = s_0 = SSS(0)$; likewise, each interpolated new share $s_i'$ can be thought of as $SSS_0'(i)$, or alternative shares of $s_0 = s$. Thus, when these $s_i'$ are interpolated again, the original $s$ is recovered. Intuitively, consider that the original secret $s$ is split into shares once, with each of those node shares being split up again. This broadcast changes which nodes hold which sub-shares of the original secret, although the underlying shared data remains the same. $SSS'_i(x)$ can have a different degree, and therefore a different reconstruction threshold $t'$, than $SSS(x)$, allowing users to add or remove recovery nodes. This secret refresh can also be configured to run at some desired interval (e.g. once per day) to protect against recovery nodes' shares being leaked over time.
\fi

\iflong
\else
\vspace{-6ex}
\fi
\begin{figure}
    \centering
    \caption{DPSS Secret Refreshing}
    \label{fig-secret-refresh-explanation}
\begin{tikzpicture}[node distance=1.2cm, font=\small]
    \node (tikzanchor) [] {};
    \node (contact1) [align=right, anchor=east, below=.5em of tikzanchor] {Node 1};
    \node (contacti) [align=right, below=.8cm of contact1.east, anchor=east] {Node $i$};
    \node (contactt) [align=right, below=.8cm of contacti.east, anchor=east] {Node $t{+}1$};
    \node (contacttanchor) [anchor=east, below=.8cm of contacti.west] {};
    \draw [decorate,decoration={brace,amplitude=5pt,mirror,raise=4ex}]
  (contact1.north west) -- (contacttanchor.south) node[midway,xshift=-2.5em,align=right, anchor=east]{Former\\Recovery\\Nodes};
    
    \node (contactj) [align=right, below=1.2cm of contactt.east, anchor=east] {New Recovery Node $j$};
    
    \node[right=8.9cm of contact1.east, anchor=east] (contact1_ground) {};
    \node[right=8.9cm of contacti.east, anchor=east] (contacti_ground) {};
    \node[right=8.9cm of contactt.east, anchor=east] (contactt_ground) {};
    \node[right=8.9cm of contactj.east, anchor=east] (contactj_ground) {};
    
    \draw[->] (contact1) -- (contact1_ground);
    \draw[->] (contacti) -- (contacti_ground);
    \draw[->] (contactt) -- (contactt_ground);

    \node (dotted) [align=right, below=.4cm of contactt.south west, anchor=east] {};
    \node[below right=.4cm and -.3cm of contactt_ground.south east] (dotted_ground) {};
    \draw[dotted] (dotted) -- (dotted_ground);
    
    \node (contact11) [right=0.5cm of contact1.east, fill=white, draw, anchor=center] {$s_1$};
    \node (contacti1) [right=0.5cm of contacti.east, fill=white, draw, anchor=center] {$s_i$};
    \node (contactt1) [right=0.5cm of contactt.east, fill=white, draw, anchor=center] {$s_{t+1}$};
    
    \node (contact12) [right=1.0cm of contact1.east, anchor=east] {};
    \node (contacti2) [right=2.2cm of contacti.east, anchor=east] {};
    \node (contactt2) [right=3.4cm of contactt.east, anchor=east] {};

    \node (contact13) [below right=2.8cm-0.1cm and 1.4cm of contact12.east] {};
    \draw[->] (contact12) -- node[midway, above, sloped, fill=white] {$SSS_1'(j)$} (contact13);
    
    \node (contacti3) [below right=2cm-0.1cm and 0.950cm of contacti2.east] {};
    \draw[->] (contacti2) -- node[midway, above, sloped, fill=white] {$SSS_i'(j)$} (contacti3);
    
    \node (contactt3) [below right=1.2cm-0.1cm and 0.50cm of contactt2.east] {};
    \draw[opacity=0,->] (contactt2) -- node[midway, above, sloped, fill=white, opacity=1] {$SSS_{t+1}'(j)$} (contactt3);
    \draw[->] (contactt2) -- (contactt3);

    \draw[->] (contactj) -- (contactj_ground);

    \node (contactjanchor) [right=1.9cm of contactt3] {};
    \node (contactjinterp) [rectangle callout, callout absolute pointer=(contactjanchor), draw, fill=white, minimum width=3cm, align=center, above=2px of contactjanchor.north, inner sep=7px, text depth=2px] {Multiply responses\\ by $\lambda_i(0)$ and sum\\$=SSS_0(j)= s'_j$ };
    
    \node (contactjs) [right=1.65cm of contactjanchor.east, draw, fill=white] {$s'_j$};
\end{tikzpicture}
\end{figure}
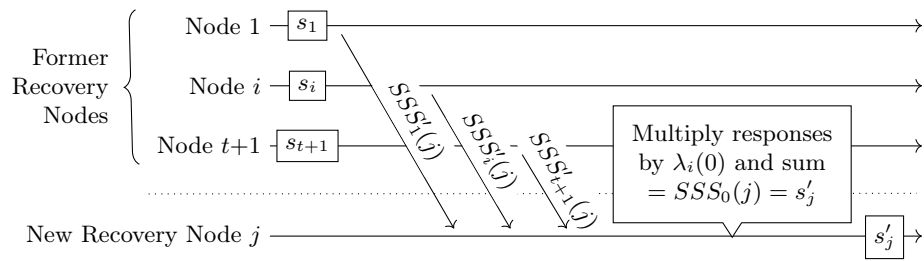

\iflong
Honey Badger is able to tolerate up to one-third of nodes being Byzantine and not following the protocol. The Kintsugi prototype currently does not implement these aspects of Honey Badger (e.g. the multi-valued validated Byzantine agreement), and hence requires all recovery nodes to correctly follow the protocol and only tolerates honest-but-curious or offline nodes. There is currently no way to determine which secret shares are valid for interpolation in order to recover $s$, which would be an issue in the presence of Byzantine recovery nodes. Kintsugi can be made Byzantine fault-tolerant in the future by implementing the remainder of the Byzantine agreement protocols.
\fi

\iflong
Honey Badger requires a designated dealer node to calculate and broadcast the initial secret shares. In Kintsugi, the user whose key is being recovered acts as the dealer. This level of trust is acceptable, because the user's key is ultimately being recovered and we can assume the user behaves honestly. Applications that wish to avoid trusted setup can use a distributed key generation (DKG) protocol, such as the one proposed by Das et al. \cite{Das_Xiang_Kokoris-Kogias_Ren_2022}.
\fi

\iflong
We also weighed other asynchronous DPSS protocols: in particular, we considered DyCAPS \cite{Hu_Zhang_Chen_Zhou_Jiang_Liu_2022}, a protocol with similar fault-tolerance guarantees and communication costs. We ultimately chose Honey Badger over DyCAPS due to Honey Badger's support of the Ristretto curve in existing OPRF instantiations and ease of development, whereas DyCAPS requires a pairing-friendly curve, such as the BLS12-381 curve, and would have required a custom OPRF design.
\fi

\subsection{Changing Recovery Nodes}

When users change their recovery nodes or their recovery threshold $t$, they initiate a DPSS refresh. Users then update their DHT entry of recovery nodes to hold the new set of recovery nodes, and the former recovery nodes destroy their old secret shares. 

\iflong
Former recovery nodes cannot participate in current key recovery attempts, thanks to the DPSS refresh protocol. Formerly valid secret shares, interpolated with current, refreshed shares, will fail to reconstruct $s$. For additional security, former recovery nodes are expected to cooperate and delete their old shares. Otherwise, it is possible for $t_{old}+1$ former recovery nodes to collude and recover $s$, for the threshold $t_{old}$ at the point when the former recovery nodes were valid. Still, colluding recovery nodes cannot gain any information about $P$ without mounting a brute-force attack.
\fi

\section{Discussion}

Kintsugi's approach provides several key benefits:

\begin{itemize}
    \item \textbf{Decentralization.}
    \iflong
    Authentication mechanisms like OPAQUE \cite{Wood_Bourdrez_Lewi_Krawczyk}, and most other E2EE platforms (see Table \ref{table:comparison}), rely on a single point-of-failure server for recovery. Distributing secret shares across multiple recovery nodes mitigates the concern that the server may mount attacks (e.g. brute-force) on the user's password or that the server secret is leaked.
    \else
    Instead of relying on a single provider for recovery, Kintsugi distributes secret shares across multiple recovery nodes. This mitigates the concern that a single party may mount attacks (e.g. brute-force) on the user's password or that the server secret is leaked.
    \fi
    \item \textbf{Recovery from lost devices.}
    Because Kintsugi is based on a password, users can access their recovery key backups on a different device, in contrast to recovery schemes requiring hard copies of data. By design, Kintsugi does not provide any means of recovery if the user loses their password.
    \item \textbf{Brute-force and collusion resistance.}
    \iflong 
    Kintsugi makes it relatively safe to use a lower-entropy password. If the backup encryption key were directly derived from the password without an OPRF exchange, offline brute-force attacks on the user's password would be feasible due to the low entropy of most passwords, whereas the OPRF output point used as an encryption key in Kintsugi has higher entropy. Kintsugi also requires the participation of recovery nodes to reconstruct the key, so attackers cannot directly perform an offline brute-force without compromising at least $t+1$ secret shares. On the other hand, the rate-limiting performed by recovery nodes protects against online brute-force attacks.
    \else
    Kintsugi makes it relatively safe to use a lower-entropy password: the secret-sharing across recovery nodes protects against offline brute-force attacks, and the rate-limiting performed by recovery nodes protects against online brute-force.
    \fi
\end{itemize}

A limitation of Kintsugi is that users must still keep track of their password. \iflong However, passwords remain a common experience that many users are comfortable with, as opposed to storing recovery codes or hard copies of recovery files. \else However, passwords remain a common experience that many users are comfortable with, as opposed to storing high-entropy recovery data. \fi \iflong Kintsugi's usage of an OPRF output based on a low-entropy password is a sweet spot between short, easy-to-remember means of recovery like PINs \cite{Connell_Signal_Key_Recovery} and high-entropy mechanisms such as Bitcoin seed phrases \cite{BIP39}. \fi

\iflong Kintsugi's concept of recovery nodes is flexible: it also allows the user's contacts' devices to serve as recovery nodes, similar to the social recovery methods used on platforms like PreVeil \cite{Preveil_Whitepaper} or Apple iCloud \cite{Apple_Recovery_Contact}. One could consider an instantiation of Kintsugi leveraging email, SMS, or chat apps instead of our use of libp2p to conduct the OPRF evaluations. The user's contacts would act as a human-enforced "social rate limit" to prevent online brute-force attacks, since contacts are unlikely to respond to high volumes of requests. \fi

\iflong
\section{Related Work}

Table \ref{table:comparison} compares the recovery mechanisms and properties of several E2EE services. Blessing et al. \cite{Blessing_Hugenroth_Anderson_Beresford_2024} have highlighted several concerns with existing E2EE platforms, including risks of total account lockout due to improper storage of recovery codes or files used by apps like WhatsApp and 1Password. As shown in the table, few platforms use decentralized recovery mechanisms, and those relying on social contacts for recovery are often vulnerable to social engineering.

\begin{sidewaystable}
    \caption{Comparison of E2EE platforms and their recovery mechanisms.}
    \label{table:comparison}
\centering
\begin{tabular}{@{}lllcccccccc@{}}
\toprule
& & & \multicolumn{5}{c}{Method}  & \multicolumn{2}{c}{Properties} \\
    \cmidrule(lr){4-8}\cmidrule(lr){9-11}
 & \begin{tabular}[t]{@{}l@{}}Service\\name\end{tabular} & \begin{tabular}[t]{@{}l@{}}Data\\recovered\end{tabular} & Password$^*$ & \begin{tabular}[t]{@{}l@{}}Recovery\\ codes\end{tabular} & PIN & \begin{tabular}[t]{@{}l@{}}Recovery\\ files\end{tabular}  & \begin{tabular}[t]{@{}l@{}}Social\\ contacts\end{tabular} & \begin{tabular}[t]{@{}l@{}}Resistant\\ to social\\engineering\end{tabular} & \begin{tabular}[t]{@{}l@{}}HSM-\\ based\end{tabular} & \begin{tabular}[t]{@{}l@{}}Decent-\\ ralized\end{tabular} \\ \midrule
\multirow{3}{*}{\begin{tabular}[t]{@{}l@{}}Messaging,\\ email\end{tabular}} & Signal & Key & - & - & \checkmark \cite{Connell_Signal_Key_Recovery} & - & - & N/A & \checkmark \cite{Connell_Signal_Key_Recovery} & - \\
 & WhatsApp & Account  & \checkmark \cite{Whatsapp_Recovery} & \checkmark \cite{Whatsapp_RecoveryCodes} & - & - & - & - & \checkmark \cite{Whatsapp_Recovery} & - \\
 & PreVeil & Key & - & - & - & - & \checkmark \cite{Preveil_Whitepaper}& N/A & - & \checkmark \cite{Preveil_Whitepaper} \\
 \hline&\\[-1.0em]
\multirow{3}{*}{File storage} & Apple iCloud & Account & - & \checkmark \cite{Apple_Recovery_Key} & - & - & \checkmark \cite{Apple_Recovery_Contact} & - & \checkmark \cite{Apple_HSM} & - \\
 & MEGA & Account & - & \checkmark \cite{MEGA_Recovery_Key} & - & - & - & N/A & - & - \\
 \hline&\\[-1.0em]
\multirow{2}{*}{\begin{tabular}[t]{@{}l@{}}Password\\ manager\end{tabular}} & 1Password & Account & - & \checkmark \cite{1Password} & - & \checkmark \cite{1Password} & \checkmark \cite{1Password} & - & - & - \\
 & LastPass & Account & \checkmark \cite{LastPass} & \checkmark \cite{LastPass} & - & - & - & N/A & - & - \\
 \hline&\\[-1.0em]
\multirow{2}{*}{\begin{tabular}[t]{@{}l@{}}Misc.\\ (not E2EE)\end{tabular}} & SSS \cite{Shamir_1979} & Key  & - & - & - & - & \checkmark \cite{Shamir_1979} & - & - & \checkmark \\
 & Bitcoin & Key & - & \checkmark \cite{BIP39} & - & - & - & - & ** & \checkmark \\
 & OPAQUE & Key & \checkmark \cite{Wood_Bourdrez_Lewi_Krawczyk} & - & - & - & - & \textsc{N/A} & - & - \\
 \hline&\\[-1.0em]
Our paper & Kintsugi & Key & \checkmark & - & - & - & \checkmark & \checkmark & - & \checkmark \\ \bottomrule&\\[-.25em]
\end{tabular}
\\
\footnotesize{$^*$ In this table, passwords refer to user-chosen, memorable secrets, whereas recovery codes are \\ automatically generated. PINs are also user-chosen, but shorter and low-entropy. \\ $^{**}$ Hardware crypto wallets act as consumer-grade HSMs.}
\end{sidewaystable}

\subsection{Social Recovery and Trust Distribution}

Several E2EE services rely on social recovery to provide some aspect of trust distribution or decentralization. For example, Apple iCloud allows users to designate a single recovery contact who can help the user regain access to their account \cite{Apple_Recovery_Contact,Blessing_Hugenroth_Anderson_Beresford_2024}. The account owner is asked some verification questions as authentication, and iCloud has a waiting period in place to alert the account owner to potentially malicious recovery attempts. Nevertheless, a trusted recovery contact is often close enough to the user to guess answers to verification questions. In addition, the failsafe timeout relies on the user being online during that period to notice any notifications. This approach is also centralized, as Apple is a required intermediary in the recovery process.

1Password, an E2EE password manager, also supports social recovery via Recovery Groups assigned to teams of users \cite{1Password_Whitepaper}. Each Recovery Group member is able to help users recover access to their account. The whitepaper notes that this recovery process should require some out-of-band verification of requests and that the onus is on Recovery Group members to avoid social engineering. As well, because each Recovery Group member is able to unilaterally recover access, a single malicious recovery contact can gain access to the user's verification requests via email access and ultimately access the user's account. This is less secure than threshold-based designs, where recovery contacts must also compromise the other contacts' secrets to reconstruct the user's recovery key.

PreVeil is another E2EE platform, focusing on email and file collaboration. \iflong Among the E2EE providers listed in Table \ref{table:comparison}, it is the only service that supports decentralized recovery. \fi PreVeil supports the notion of Approval Groups, a threshold social key recovery scheme, but their whitepaper does not explain how these groups protect against approvers colluding to recover the key \cite{Preveil_Whitepaper}. PreVeil also supports Express Account Recovery, a feature that requires two shares to recover the key, with the user storing one share and the other being stored on the PreVeil server \cite{Preveil_Product_Release}. However, this requires the user to safely store their share in an accessible location and relies on PreVeil as a single trusted party in recovering the user's key.

Note that none of the E2EE platforms listed in Table \ref{table:comparison} that support social recovery contacts are resistant to social engineering attacks. An adversary may trick social contacts into providing their secret shares or initiating a recovery operation. Alternatively, social contacts may collude to gain access to the user's account. Some prior work on decentralized key recovery provides protection against individual curious nodes, but not gossiping, honest-but-curious nodes \cite{Anderson_Stajano_2014}.

\subsection{OPAQUE}

We drew inspiration from the password-authenticated key exchange (PAKE) protocol OPAQUE \cite{Wood_Bourdrez_Lewi_Krawczyk} for Kintsugi. OPAQUE provides a way for users to authenticate themselves to a server without revealing their password to the server in plaintext. OPAQUE specifies methods for registration and login: in Kintsugi's design, we reframe login as key recovery while applying similar OPRF flows. Also, in OPAQUE, a single malicious server can perform an offline brute-force attack to guess the user's password; in Kintsugi, a threshold of recovery nodes must collude before offline brute-force is possible.

\subsection{Threshold OPRFs}

Several other threshold OPRF-based systems exist. For instance, Jarecki et al. propose an updatable, oblivious key management system for encrypted storage systems \cite{Jarecki_Krawczyk_Resch_2019}, based on their prior threshold OPRF work \cite{Jarecki_Kiayias_Krawczyk_Xu_2017}. Their work describes a key management service with which clients perform an OPRF exchange to encrypt data and its decentralized, threshold OPRF-based variant. Although their system utilizes proactive secret sharing, the system does not specify how recovery nodes can be dynamically changed. It also requires the joint, distributed generation of new secrets and a distributed multiplication protocol, whereas Kintsugi avoids the need for any distributed key generation by relying on the user to deal secret shares. However, their system supports rotation of the OPRF secret $s$ such that the encrypted data can be exclusively decrypted by new keys, while Kintsugi only rotates the secret shares and maintains the same shared $s$. 

Juicebox is another decentralized key recovery protocol, based on PIN authentication and threshold OPRF exchanges \cite{Juicebox}. Its design distributes trust across "realms", representing independent service providers. Juicebox does not allow updating these realms or the user's recovery threshold $t$ after registration, and it requires at least some of these realms to be HSM-based in order to rate-limit PIN guesses. Interestingly, it also allows for HSM-based realms to be supplemented by software-backed providers, as long as a sufficient threshold of realms is reached in total. This provides a more cost-effective and scalable approach, compared to other methods which are entirely reliant on HSMs, like WhatsApp's and Signal's recovery schemes.

\fi

\section{Conclusion}

Decentralized recovery mechanisms for E2EE services mitigate the risks of typical, centralized recovery flows. Relying on trusted hardware under a single provider's control can pose concerns for applications requiring metadata privacy or lacking financial resources. In this paper, we propose Kintsugi, a decentralized recovery protocol that distributes trust over multiple recovery nodes. Future work may include implementing Byzantine fault-tolerance and exploring alternative instantiations, such as a PKI-based recovery flow instead of relying on rate-limiting. In addition, we plan to integrate Kintsugi as an extension module for the Automerge CRDT library as part of a project to support authentication \cite{Automerge,Beehive}. Overall, Kintsugi offers a new outlook on secure recovery protocols, eliminating the need for centralized infrastructure while allowing users to recover from device loss and maintaining strong security properties.

\begin{credits}
\subsubsection{\ackname} Emilie Ma conducted this work as a visiting researcher at the University of Cambridge.

\subsubsection{\discintname}
The authors have no competing interests to declare that are
relevant to the content of this article. 
\end{credits}

\bibliographystyle{splncs04}
\bibliography{bibliography}

\end{document}

\speaker{William Wayman} When the user is setting out the nodes that they want to share the secrets with, is there a central trusted party that has knowledge of these nodes?

\reply The user has knowledge of the nodes, and this is then replicated in a distributed hash table.

\speaker{William Wayman} So the user needs to remember these nodes?

\reply No, this DHT is replicated across all of the Kintsugi nodes. The user can access these nodes later by querying any recovery node.

\speaker{William Wayman} So there's no central party that knows the nodes that a given user wants to access. There's no representation of that.

\reply Yes, there's no central authority because this list is distributed across all other nodes. The distributed hash table holds these mappings.\footnote{We updated the paper to additionally clarify that there is currently no confidentiality nor integrity guarantee for recovery node information.}

% Maybe what William is trying to get at (but doesn't quite say): there is no confidentiality nor integrity guarantee for the recovery node information -- that is, anybody can look up who a user's recovery nodes are, and a misbehaving DHT node could serve up fake recovery node information. If the recovery nodes were stored on a central server, we'd perhaps get a simpler threat model than with a DHT that anybody can join. Maybe it would be worth addressing that concern explicitly in the paper?

\speaker{Fabio Massacci} I like all your examples, but what is the human's interface in your protocol? All of your first examples were a human that has lost their device. Here, you're instead making an example where the user needs a computation device that can compute an OPRF, so you need another device to actually recover the keys and do all these other things.

\reply Recovery takes place locally on the user's device.

\speaker{Fabio Massacci} So you need a second device to actually recover the first device that has been lost.

\reply Yes, there needs to be recovery servers or other end user devices registered as recovery nodes. We assume the user has just bought a new device that they want to recover their data onto and that the recovery flow and its OPRF computation will take place on this new device. 

\speaker{Jonathan Anderson} If you're not requiring the recovery nodes to do authentication, then if those nodes are being used a lot, how do you do the rate-limiting? If there's 10,000 users who are using recovery node 3, what kind of rate-limiting does it do?

\reply The rate-limiting is done based on the purported user that's attempting recovery, which is limited locally on that recovery node to once every X seconds or so.

\speaker{Jonathan Anderson} That means an attacker could perform a denial of service attack against legitimate recovery attempts by pretending to be me, invoking the rate-limiting, but they wouldn't be able to make lots of password guesses. Is that right?

\reply Yes, I believe so.\footnote{We are actively brainstorming how to more robustly perform rate-limiting to also protect against DoS, and the paper proposes IP-based rate-limiting as an alternative that comes with its own tradeoffs. In the meantime, however, we argue that genuine recovery attempts occur so infrequently that this is not a major concern.}

\speaker{Fabio Massacci} So, as a recovery node, you need to know the identity of the user, because otherwise you'll never be able to put a rate limit on them.

\reply You just need to know the purported identity. User Y could be attempting to recover User X, for example. 

\speaker{Mark Lomas} So if I've understood this correctly, the user knows that they've succeeded in running this protocol, because they've got their data back. But does anyone other than the user know that it succeeded?

\reply No.

\speaker{Mark Lomas} Doesn't that worry you? Wouldn't it be useful to have an audit trail that says "Somebody appears to be trying to break into Mark Lomas' account"? We don't know whether it was actually Mark's device, or Susan breaking in. 

\reply I would imagine it would be easy to have an audit log that tracks recovery attempts made by purported users.

\speaker{Mark Lomas} But you won't know the difference between a successful and a failed attempt.

\reply But as a user, you'd be able to see, oh, this wasn't me. 

\speaker{Mark Lomas} The user could know, but none of the other components in the system would know. 

\speaker{Martin Kleppmann} (Co-author of paper.) If you wanted the recovery nodes to be able to find out whether a recovery attempt was successful, you could always have the user, after they decrypted the envelope, then send back a signed statement to the recovery notes to let them know after the fact that it was successful. And if no such statement arrives, then you assume that it was unsuccessful. 

\speaker{Fabio Massacci} I think the problem that Mark wants to solve is how someone else, not the user, can know if somebody's trying to forge the user's password. You can't assume the user is cooperating if no one's aware.

\speaker{Tyler Moore} What would the advantages or disadvantages of adopting the system be, from the perspective of the platforms themselves? I can see how this has better security properties than existing efforts. That's good. But is there any advantage or disadvantage to say, Signal, adopting this versus their current protocol? It sounds like these recovery nodes could operate this decentralized infrastructure, or are you envisioning that the recovery nodes would be operated by the users themselves? What are the kind of the costs of operating this, compared to the current approaches? I'm just wondering if that could be an impediment to these services wanting to adopt your proposal.

\reply The existing services already have a different type of decentralized infrastructure based on having all these HSMs, even if they're run on different cloud providers, under one company's control, as Signal does. We have to use decentralized recovery nodes, so you might have to spin up extra servers. As for the cost of running each of these recovery nodes, you can either use end user devices, so it could be that one Kintsugi user is acting as a recovery node for another Kintsugi user, or users can run their own recovery servers, sort of like Tor's hobbyist node operator model. These servers can all be operated by different service providers, which is a main advantage of the whole decentralization idea.

\speaker{Liqun Chen} Is it true that under the secret sharing scheme, anyone else can request recovery but cannot get the right message back because they're not the original user? Does this mean the secret that the user shares cannot be a low-entropy password? Should it be a high entropy secret?

\reply Due to the threshold OPRF construction, it doesn't matter what the user provides as their password, which becomes the OPRF secret. It's okay to use a lower entropy password because under the hood, the OPRF multiplies the secret by a random scalar, which provides more entropy. Then, assuming the discrete log problem, this becomes safe to combine with the recovery nodes' secret, so entropy is less of a problem. 

\speaker{Frank Stajano} At some point, you said that this method was impervious to social engineering. I am constitutionally skeptical about this type of claim. Could you please elaborate?

\reply I said this in the context of existing social recovery schemes, where you tell your friend, "Hi, I lost my phone. Can you help get me back in?" Your friend has to decide if this is the genuine person that they know. With our scheme, because the recovery nodes operate blind to whatever the user has blinded their password to be, the recovery nodes don't have to make this judgment call on whether or not a recovery request was authentic.

\speaker{Frank Stajano} On what basis does a recovery node make its decision to cooperate and apply its share? 

\reply It will always cooperate.\footnote{We assume no Byzantine nodes. Byzantine fault-tolerance can be integrated into this protocol as a result of other protocol building blocks, but has not yet been implemented.}

\speaker{Martin Kleppmann} The one thing that you have to trust the recovery nodes to do is to perform the rate limiting, so to not allow more than, say, one password guess per second per username. 

\speaker{Frank Stajano} So the claim could be rephrased as saying ``It is totally socially engineerable, in the sense that they will always respond, but because there's a limit, they can't go very far.''

\speaker{Martin Kleppmann} Yes, and with every communication with the recovery nodes, the adversary can only try one password guess, whereas with a social recovery system, if you manage to convince the recovery contact once, then you're in.

\speaker{Frank Stajano} So ``impervious to social engineering'' could have meant ``It is not possible to persuade the recovery nodes to cooperate. Instead they cooperate all the time, but we have a rate limit, and therefore it does not matter.''

\speaker{Tyler Moore} I think it means that her protocol does not rely on a human making a critical decision, which is the component which is inherently socially engineerable.

\speaker{Tyler Moore} Last question: why do you call it Kintsugi?

\reply{Kintsugi is the Japanese art of mending together broken pottery with gold. One would fill in the cracks in a broken vase, for example, with gold in order to highlight the imperfections while making the vase better and new again. In this case, we're mending together our distributed key shares into this stronger, decentralized key recovery protocol. Thank you.